\documentclass[a4paper,onecolumn]{rsauthor}
\usepackage{amsmath}
\usepackage{amssymb}
\usepackage{amsfonts}
\usepackage{amsthm}
\usepackage{graphicx}
\usepackage{endnotes}
\usepackage{setspace}
\usepackage{verbatim}
\usepackage{times}
\usepackage{helvet}
\usepackage{courier}
\usepackage{bm}
\usepackage{url}
\usepackage[english]{babel}
\usepackage{dcolumn}
\usepackage{color}
\usepackage{geometry}
\usepackage{textcomp}
\usepackage{subfigure}
\usepackage[round]{natbib}
\usepackage{xspace}
\usepackage{stfloats}
\usepackage{balance}
\usepackage{lastpage}
\usepackage{rotating}
\usepackage{latexsym}
\usepackage{multicol}
\usepackage{array}
\usepackage{algorithm}

\graphicspath{{figures/}{}}

\renewcommand{\d}{\mathrm{d}}

\title{The bending of an elastic beam by a liquid drop: A variational approach}

\author{S\'ebastien Neukirch$^{1,2}$, Arnaud Antkowiak$^{1,2}$, \& Jean-Jacques Marigo$^3$}

\address{
~$^1$ CNRS, UMR 7190, Institut Jean Le Rond d'Alembert, F-75005 Paris, France.\\
~$^2$ UPMC Univ Paris 06, UMR 7190, Institut Jean Le Rond d'Alembert, F-75005 Paris, France.\\
~$^3$ CNRS, Ecole Polytechnique, UMR 7649, Lab. M\'eca. Solides, F-91128 Palaiseau Cedex, France.
}

\date{\today}

\abstract{
We study the interaction of a liquid drop with an elastic beam in the case where bending effects dominate. We use a variational approach to derive equilibrium equations for the system in the presence of gravity and in the presence or absence of contact line pinning. We show that the derived equilibrium equations for the beam subsystem reveal the external forces applied on the beam by the liquid and vapor phases. Among these, the force applied at the triple line (the curve where the three phases meet) is found to lie along the liquid-vapor interface.
}
\begin{document}

\maketitle

%===========================================%

\section{Introduction}
%==============
%
%
%In 1959, Richard Feynam gave a lecture at an American Physical Society meeting at Caltech entitled `There's Plenty of Room at the Bottom', where he suggested that tiny machines could directly work at the nanoscale and even directly manipulate atoms.
%
The present trend toward miniaturization of engineering systems and machines is giving surface effects the leading role: in a system of size $L$ the respective scalings of volume ($L^3$) and surface ($L^2$) forces are such that the latter outrange the former as soon as $L$ is small enough.
Among other surface effects, surface tension is now widely used at small scales, for example to self-assemble microsystems \citep{syms2003}.
The concept of force is not easy to explain to recalcitrant students: Has anyone already seen a force? How to be sure of the direction of an applied force? Capillary forces are no exception and conceptual questions about it recurrently emerge \citep{Marchand-Weijs:Why-is-surface-tension-a-force:2011}. As these capillary forces are now used to bend small elastic structures \citep{roman2010} and as it has been recently proposed that these forces might not act as previously thought \citep{Marchand-Das:Capillary-Pressure-and-Contact:2012}, we here study the interaction of a liquid drop with a flexible beam from an energy point of view: we derive equilibrium equations of the system from a variational approach that is merely built on the classical hypothesis of the presence of surface energies arising at interfaces between the three phases: solid, liquid, and vapor.

In section \ref{section:drop-on-rigid} we recall that the Young-Dupr\'e relation for the contact angle of a drop lying on a substrate can be derived from a variational approach where the concept of force is not invoked, as first realized by \citet{Gauss:Principia-Generalia-Theoriae:1830,Gauss:Werke:1877}. In section \ref{section:flex-no-g} we consider the case where the substrate is a flexible beam and we add gravity (for the beam and the liquid) in Section \ref{section:with-gravity}. We recall in Section \ref{section:contact-line-pinning} that contact line pinning can also be considered from an energy point of view, and we finally illustrate our model by computing the behavior of a drop-beam system as the drop evaporates. Conclusion follows in Section \ref{section:conclusion}.

%===========================================%
\section{Liquid drop on rigid substrate, no gravity} \label{section:drop-on-rigid}
\begin{figure}[thb]
\centering
\includegraphics[width=0.99\columnwidth]{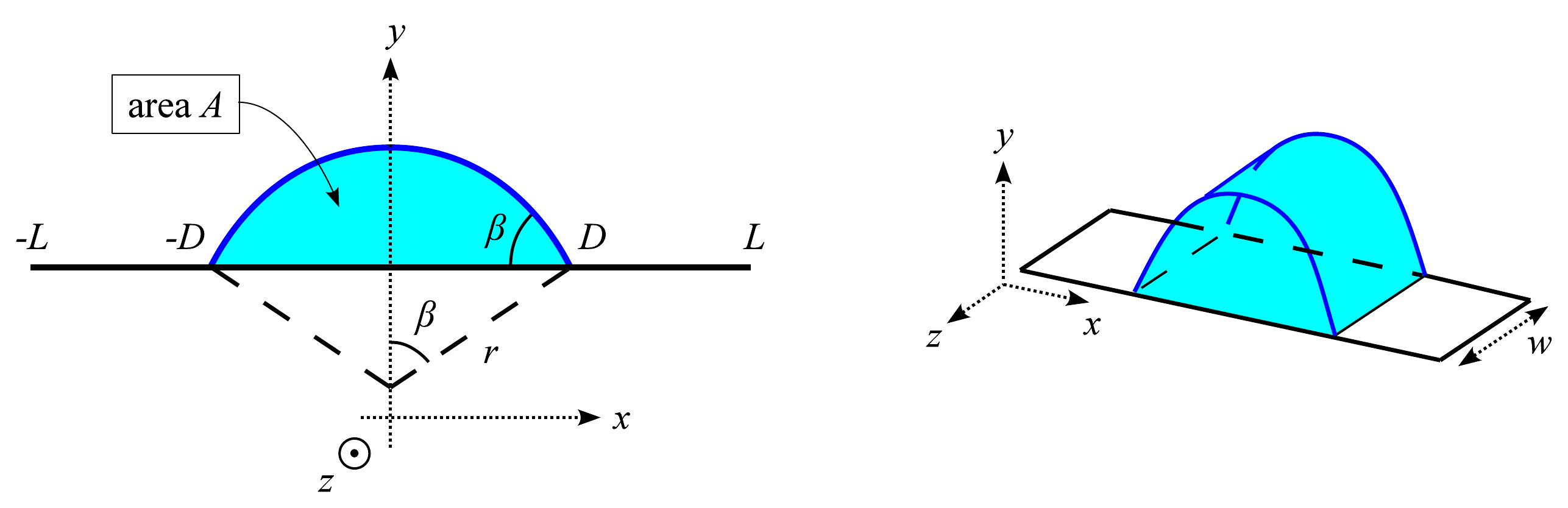}
\caption{A two-dimensional model of a liquid drop at rest on a rigid substrate. In the absence of gravity the liquid-vapor interface is a circular arc of radius $r$. The contact angle $\beta$ is set by the balance of the interfaces energies, see Eq.~\ref{equa:laplace-pressure}, and the wetted length $2D$ depends on the drop volume $V$.}
\label{fig:young-dupre}
\end{figure}

%==================
%
%
We consider the equilibrium of a liquid drop of given volume sitting on a rigid substrate of length $L$, see Fig. \ref{fig:young-dupre}. If the drop is small enough, gravity and the hydrostatic part of the pressure can be neglected and consequently the liquid-air interface is circular.
For simplicity we adopt the two dimensional framework introduced in \cite{Rivetti-Neukirch:Instabilities-in-a-drop-strip-system::2012} where the liquid-vapor interface is a cylindrical arc; We call $r$ its radius, $w$ its height, and $2\beta$ its opening angle. The liquid vapor interface then comprises $(i)$ a cylindrical surface of area $2 \beta r w$ and $(ii)$ two planar caps, each of area $A=r^2 (\beta - \sin \beta \, \cos \beta)$. 
The wetting angle is equal to $\beta$ and the wetted length of the beam is noted $2D$.
To each of the three different interfaces, liquid-solid, liquid-vapor, and solid-vapor we associate an energy per area:  $\gamma_{\ell s}$ , $\gamma_{\ell v}$, and  $\gamma_{sv}$ respectively. 
The energy of the system is then given by the sum:
\begin{equation}
E(\beta, r,D)=2 (w\beta r+ A )  \gamma_{\ell v} + 2 w D \gamma_{\ell s} + 2 w (2L-D) \gamma_{sv}
\label{equa:interface-energy}
\end{equation}
To minimize $E$ under the constraints of $(i)$ fixed volume $V=Aw=w r^2 (\beta - \sin \beta \, \cos \beta)$ and $(ii)$ geometric relation $D=r \sin \beta$, we then introduce the Lagrangian:
\begin{equation}
{\cal L}=E- \eta w (2 D - 2 r \sin \beta) - \mu w \left[ r^2 (\beta - \sin \beta \, \cos \beta) \right]
\end{equation}
where $\eta$ and $\mu$ are Lagrange multipliers.
We note $\gamma=\gamma_{\ell v}$ and $\Delta \gamma=\gamma_{\ell s} - \gamma_{sv}$.
Equilibrium equations are found by imposing that derivatives of ${\cal L}$ with regard to the three variables $\beta, r,D$ vanish:
\begin{subequations}
\begin{align}
\frac{\partial {\cal L}}{\partial \beta} = &0 = 2 w r \gamma + 2 \eta w r \cos \beta - \mu w r^2 (1-\cos 2 \beta) \label{equa:YD-eq1}\\
\frac{\partial {\cal L}}{\partial r} = &0 = 2 w \beta \gamma + 2 \eta w \sin \beta - 2 \mu w r (\beta - \sin \beta \cos \beta) \label{equa:YD-eq2}\\
\frac{\partial {\cal L}}{\partial D} = &0 = 2 w \Delta \gamma - 2 \eta w \label{equa:YD-eq3}
\end{align}
\end{subequations}
Combining (\ref{equa:YD-eq1}) $\cos \beta$ + (\ref{equa:YD-eq2}) $r \, \sin \beta$ and (\ref{equa:YD-eq1}) $\sin \beta$ - (\ref{equa:YD-eq2}) $r \, \cos \beta$ yields $\gamma = \mu r$ and $\gamma \cos \beta + \eta=0$, and using (\ref{equa:YD-eq3}) gives $\eta=\Delta \gamma$. We finally arrive at:
\begin{align}
\Delta \gamma + \gamma \cos \beta =& 0 \label{equa:YD-relation}\\
\mu =& \frac{\gamma}{r} \label{equa:laplace-pressure}
\end{align}
The first equation is the well-known Young-Dupr\'e relation giving the contact angle and can be interpreted as a force balance of the triple point in the horizontal direction. The second equation gives the Laplace pressure inside the liquid drop.
In the vertical direction force balance is also achieved: the vertical forces acting on the rigid substrate are the distributed Laplace pressure $\mu$ and the surface tension $\gamma_{\ell v}$, with the total downward force being $2 D w \mu$ and the total upward force being $2 \gamma_{\ell v} \sin \beta$. Using (\ref{equa:laplace-pressure}) and $D=r \sin \beta$, we see that these two forces equilibrate. We shall see in the next section that when the substrate is a thin elastic strip,  these forces induce flexural deformations.

%
%===========================================%
%
%
\section{Liquid drop on a flexible beam, no gravity} \label{section:flex-no-g}
%==============
%
%
We now consider the case of a liquid drop sitting on an elastic strip, see Fig.~\ref{fig:bent-strip}, and we look for equilibrium equations governing the bending of the elastic strip by capillary forces. We still work under the hypothesis where gravity and the hydrostatic part of the pressure can be neglected, yielding  a circular liquid-air interface. In addition to the sum of the three interface energies:
\begin{equation}
E_{\gamma}=2(w\beta r+ A ) \gamma_{\ell v} + 2 w D \gamma_{\ell s} + 2 w (2L-D) \gamma_{sv}
\end{equation}
we consider the bending energy of the elastic strip.
We use the arc-length $s$ along the strip to parametrize its current position $\bm{r}(s)=(x(s),y(s))$. The unit tangent, $\bm{t}(s)=\d\bm{r}/\d s$, makes an angle $\theta(s)$ with the horizontal axis: $\bm{t}=(\cos \theta(s), \sin \theta(s))$. The bending energy density is proportional to the square of the curvature $\theta'(s)$:
\begin{figure}[ht]
    \centering
    \includegraphics[width=0.75\columnwidth]{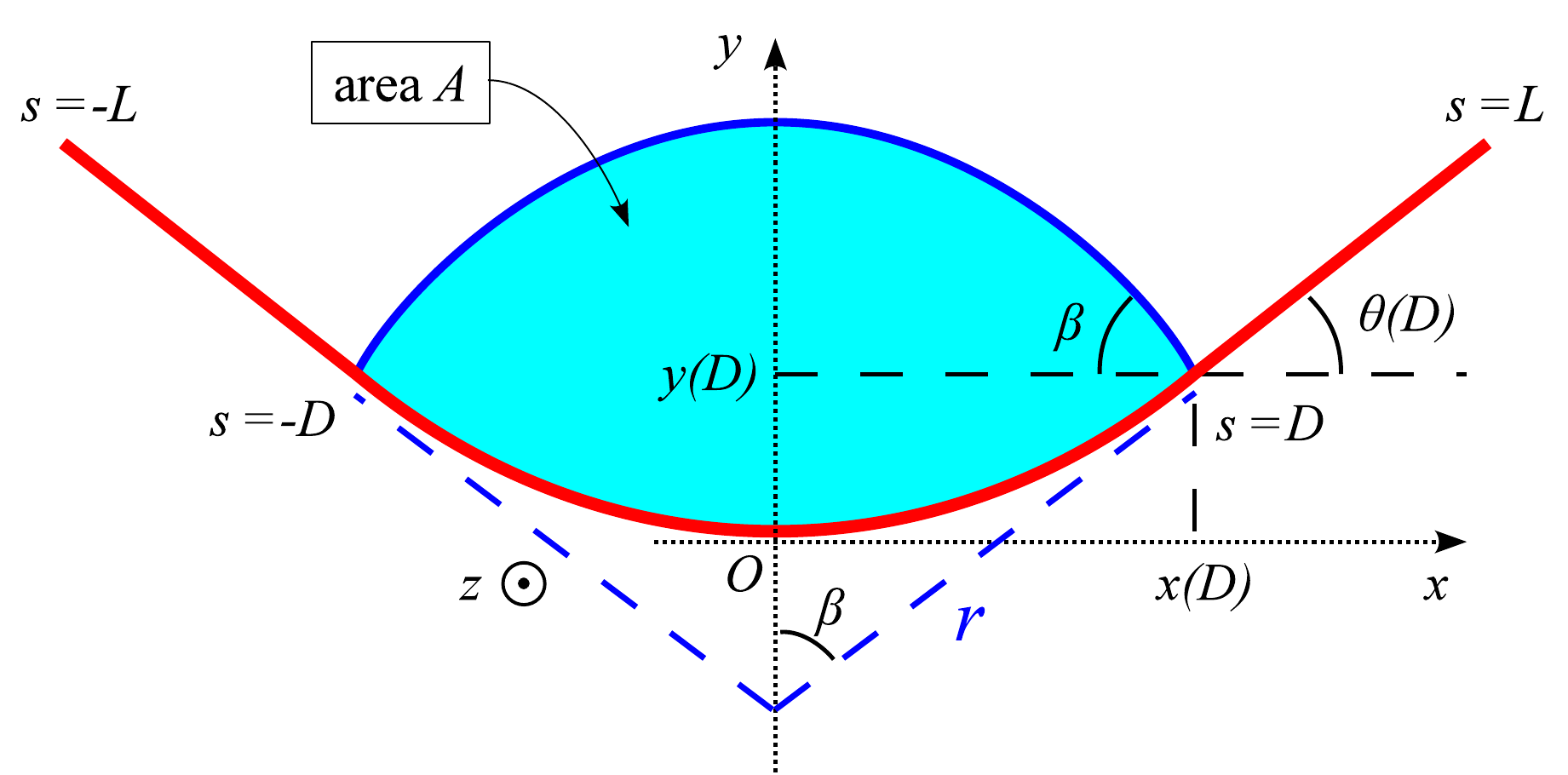}
    \caption{An elastic strip bent by capillary forces. In the absence of gravity the liquid-vapor interface is a circular arc and the strip outside the interaction region is straight. The system is invariant is the $z$ direction, with width $w$.}
    \label{fig:bent-strip}
\end{figure}
\begin{equation}
E_{\kappa} = \frac{1}{2} YI \int_{-L}^{L} \left[ \theta'(s) \right]^2 ds
\end{equation}
where $YI$ is the bending rigidity of the strip ($Y$ is Young's modulus of the beam material, and $I=h^3 w / 12$ is the second moment of area of the section of the beam).
We minimize $E=E_{\kappa}+E_{\gamma}$ under the following four constraints. First, the liquid volume $V$ is fixed. It is given by $V = w A$, where $A$ is the area in between the liquid-air interface and the liquid-solid interface:
\begin{equation}
A = r^2 \left(\beta - \frac{1}{2} \sin 2 \beta \right) + 2 \, x(D) \, y(D) - \int_{x(-D)}^{x(D)} y \, dx \label{equa:fixed_volume}
\end{equation}
Second, we have the geometric constraint:
\begin{equation}
x(D) = r \sin \beta \label{equa:xD}
\end{equation}
that is due to the intersection of the circular liquid-vapor interface and the beam at $s=D$.
As the variables $x(s)$, $y(s)$, and $\theta(s)$ all appear in the present formulation, we have to consider the continuous constraints relating them. These are our third and fourth constraints:
%And third, we have two continuous constraints relating variables $x(s)$ and $y(s)$, and the angle $\theta(s)$ as they all appear in the present formulation:
\begin{equation}
x'(s) = \cos \theta(s) \, , \; \;   y'(s) = \sin \theta(s) \label{equa:rprime}
\end{equation}
These continuous constraints necessitate the use of varying Lagrange multipliers $\nu(s)$ and $\lambda(s)$.
We therefore introduce the Lagrangian:
\begin{align}
{\cal L} =& E_\kappa + E_\gamma - \mu w \left[ r^2 \left(\beta - \frac{1}{2} \sin 2 \beta \right) + 2 \, x(D) \, y(D) - \int_{-D}^{D} y \, x' \, \d s \right]  
 \nonumber \\
& - \eta w \left[ x(D) - r \sin \beta \right] 
+ \int_{-L}^{L} \nu(s) \, \left[ x'-\cos \theta \right] \d s 
+ \int_{-L}^{L} \lambda(s) \,  \left[ y'-\sin \theta \right] \d s 
\label{equa:lagrangian1}
\end{align}
As we will only treat symmetric solutions, we focus on the positive $s$ interval: $s \in [0;L]$ with the following boundary conditions:
\begin{equation}
x(0)=0 \, , \quad y(0)=0 \, , \quad \theta(0)=0 \label{equa:BC-s0}
\end{equation}
We also remark that in this elastica model, the varying Lagrange multiplier 
$\nu(s)$ and $\lambda(s)$ will be found to be the internal force components, see Appendix \ref{appendix:elastica}. As the external force coming for the meniscus will make the internal force discontinuous as $s$ passes through $s=D$, we split the two last integrals in the Lagrangian (\ref{equa:lagrangian1}) and write:
\begin{align}
s \in [0;D) \, :& \;  \nu(s) = \nu_i(s) \, , \;  \lambda(s) = \lambda_i(s) \\
s \in (D;L] \, :& \;  \nu(s) = \nu_e(s) \, , \;  \lambda(s) = \lambda_e(s)
\end{align}
Dropping out constant terms, we arrive at:
\begin{align}
{\cal L}(x,y,\theta,\beta,r,D) &=
\frac{1}{2} YI \int_{0}^{L} \left[ \theta'(s) \right]^2 \d s   
+ w \beta r \gamma +  w D \Delta \gamma 
- \eta w \left[ \int_0^D x' \d s - r \sin \beta \right] \nonumber \\
&- \mu w \left[ \frac{r^2}{2} \left(\beta - \frac{1}{2} \sin 2 \beta \right) + \int_0^D x' \d s \, \times \,  \int_0^D y' \d s - \int_{0}^{D} y \, x' \, \d s \right]  \nonumber \\
%& \nonumber \\
&  + \int_{0}^{D} \nu_i(s) \,  \left[x'-\cos \theta \right] \d s + \int_{D}^{L} \nu_e(s) \,  \left[x'-\cos \theta \right] \d s \nonumber \\
& + \int_{0}^{D} \lambda_i(s) \,  \left[y'-\sin \theta \right] \d s+ \int_{D}^{L} \lambda_e(s) \,  \left[y'-\sin \theta \right] \d s 
\label{equa:lagrangian2}
\end{align}
where $\gamma=\gamma_{\ell v}$ and $\Delta \gamma = \gamma_{\ell s} -\gamma_{s v}$.

\subsection*{First variation}
%--------------------------------
%
The energy $E$ and the Lagrangian ${\cal L}$ are functions of the variables $x,y,\theta,\beta,r,D$. We note $X=(x,y,\theta,\beta,r,D)$ and
we consider the conditions for the state $X_e$ to minimize the energy $E$. Calculus of variation shows that a necessary condition is:
\begin{equation}
{\cal L}'(X_e)(\bar{X}) = \left. \frac{d}{d \epsilon}  {\cal L}(X_e+\epsilon \bar{X}) \right|_{\epsilon=0} \, = 0 \label{equa:1st-variation}
\end{equation}
where $\bar{X}=(\bar{x},\bar{y},\bar{\theta},\bar{\beta},\bar{r},\bar{D})$.
Moreover boundary conditions (\ref{equa:BC-s0}) implies that $\bar{x}(0)=0$, $\bar{y}(0)=0$, $\bar{\theta}(0)=0$.
Noting that:
\begin{equation}
\int_0^{A+\epsilon \bar{A}} f(x) dx = \int_0^{A} f(x) dx + \epsilon \bar{A} f(A) + O(\epsilon^2) \, 
\end{equation}
we evaluate the first variation (\ref{equa:1st-variation}) to be: 
\begin{align}
{\cal L}'(X_e)&(\bar{X}) 
= YI \int_{0}^{L} \theta' \, \bar{\theta}' \d s   
+ w \beta \bar{r} \gamma + w \bar{\beta} r \gamma 
+  w \bar{D} \Delta \gamma \nonumber \\
&
- \mu w \left[ r \bar{r} \left(\beta - \frac{1}{2} \sin 2 \beta \right) 
+ \frac{\bar{\beta} r^2}{2}  \left(1 - \cos 2 \beta \right)
\right] \nonumber \\
&- \mu w \left[ \int_0^D \bar{x}' \d s \, \times \,  \int_0^D y' \d s  + \int_0^D x' \d s \, \times \,  \int_0^D \bar{y}' \d s + \bar{D} x'(D) y(D) + \bar{D} x(D) y'(D)  \right]  \nonumber \\
&+ \mu w \left[  \int_{0}^{D} \bar{y} \, x' \, \d s +  \int_{0}^{D} y \, \bar{x}' \, \d s 
+ \bar{D} y(D) x'(D) \right]  \nonumber \\
&
- \eta w \left[ \int_0^D \bar{x}' \d s +\bar{D} x'(D) - \bar{r} \sin \beta - r \bar{\beta} \cos \beta \right] \nonumber \\
&  + \int_{0}^{D} \nu_i(s) \,  \left[\bar{x}'+ \bar{\theta}\sin \theta \right] \d s + \int_{D}^{L} \nu_e(s) \,  \left[\bar{x}'+ \bar{\theta}\sin \theta \right] \d s \nonumber \\
& + \int_{0}^{D} \lambda_i(s) \,  \left[\bar{y}'-\bar{\theta}\cos \theta \right] \d s+ \int_{D}^{L} \lambda_e(s) \,  \left[\bar{y}'-\bar{\theta}\cos \theta \right] \d s  %\nonumber \\
%& + \bar{D} [\nu_i(D)-\nu_e(D)] \, [x'(D)-\cos \theta(D)]+ \bar{D} [\lambda_i(D)-\lambda_e(D)] \, [y'(D)-\sin \theta(D)]
\label{equa:1st-variation-expression}
\end{align}
where we have used (\ref{equa:rprime}) at $s=D$ to eliminate some terms related to the last four integrals.
We require this expression to vanish for all $\bar{x}(s)$, $\bar{y}(s)$, $\bar{\theta}(s)$, $\bar{\beta}$, $\bar{r}$, and $\bar{D}$.  
For (\ref{equa:1st-variation-expression}) to vanish for all $\bar{\beta}$, we must have, as before:
\begin{equation}
r w \gamma - \mu w \frac{r^2}{2}  \left(1 - \cos 2 \beta \right)
+ \eta w r \cos \beta  = 0 \label{equa:forallbeta}
\end{equation}
For (\ref{equa:1st-variation-expression}) to vanish for all $\bar{r}$, we must have, as before:
\begin{equation}
\beta w \gamma - \mu w r  \left(\beta - \sin \beta \cos \beta \right)
+ \eta w \sin \beta  = 0 \label{equa:forallrho}
\end{equation}
Combining these last two equations we obtain:
\begin{equation}
\mu r = \gamma \text{ ~ and ~ } \gamma \cos \beta + \eta = 0
\label{equa:rho_et_eta}
\end{equation}
where $\mu$ is identified to the Laplace pressure.
For (\ref{equa:1st-variation-expression}) to vanish for all $\bar{D}$, we must have:
\begin{equation}
w \Delta \gamma - \mu w x(D) y'(D) - \eta w x'(D) = 0
\end{equation}
Using (\ref{equa:xD}), (\ref{equa:rprime}), and (\ref{equa:rho_et_eta}) we obtain:
\begin{equation}
\Delta \gamma + \gamma \cos \left[ \beta + \theta(D) \right] = 0 \label{equa:YD}
\end{equation}
This is the Young-Dupr\'e relation for the wetting angle $\beta + \theta(D)$ between the beam and the liquid-air meniscus.
Requiring (\ref{equa:1st-variation-expression}) to vanish for all $\bar{\theta}$ yields, after integration by parts:
\begin{align}
YI  \left[ \theta' \, \bar{\theta} \, \right]_0^D 
+&\int_0^D \left[ - YI \theta'' + \nu_i \sin \theta - \lambda_i \cos \theta \right] \bar{\theta} \d s  \nonumber\\
+& YI \left[ \theta' \, \bar{\theta} \, \right]_D^L
+\int_D^L \left[ - YI \theta'' + \nu_e \sin \theta - \lambda_e \cos \theta \right] \bar{\theta} \d s
=0
\end{align}
which implies that the curvature $\theta'(s)$ is continuous as $s$ goes through $D$ and that it vanishes at the $s=L$ extremity. Moreover we obtain the moment equilibrium equations:
\begin{align}
YI \theta'' =& \nu_i \sin \theta - \lambda_i \cos \theta \,  \text{ for } s \in [0;D) 
\label{equa:equilibre_moment_inside}\\
YI \theta'' =& \nu_e \sin \theta - \lambda_e \cos \theta \,  \text{ for } s \in (D;L] 
\label{equa:equilibre_moment_outside}
\end{align}
where we see that the continuous Lagrange multipliers $\nu(s)$, and $\lambda(s)$ can be identified as the $x$ and $y$ components of the internal force: $n_x \equiv \nu$ and $n_y \equiv\lambda$.
Requiring (\ref{equa:1st-variation-expression}) to vanish for all $\bar{x}$ yields, after integration by parts:
\begin{align}
\Bigl[ \bigl( - \mu w \, \left(y(D)-y \right) - \eta w + \nu_i \bigr) \, \bar{x} \, \Bigr]_0^D  %&
- \int_0^D \left( \mu w y' + \nu_i' \right) \bar{x} \, ds %\nonumber \\
%&
+ \Bigl[ \nu_e \, \bar{x} \, \Bigr]_D^L -\int_D^L  \nu_e' \, \bar{x} \, ds =0
\end{align}
The fact that we have $\bar{x}(0)=0$, but arbitrary $\bar{x}(D)$ and $\bar{x}(L)$ implies:
\begin{align}
\nu_e(L) = 0 \, , \; \;  \nu_e(D) - \nu_i(D) = - \eta w \, , \; \;
\nu'_e = 0 \, , \; \;  \nu'_i = - \mu w y' \label{equa:equilibre_force_x}
\end{align}
Requiring (\ref{equa:1st-variation-expression}) to vanish for all $\bar{y}$ similarly implies:
\begin{align}
\lambda_e(L) = 0 \, , \; \;  \lambda_e(D) - \lambda_i(D) = - \mu w x(D) \, , \; \; \lambda'_e = 0 \, , \; \;  \lambda'_i =  \mu w x' \label{equa:equilibre_force_y}
\end{align}
where we see that Laplace pressure generates an outward normal force $\mu w (y',-x')$ that causes the internal force $(\nu_i,\lambda_i)$ to vary.
In addition we see that the internal force vanishes at the $s=L$ extremity and that it experiences a discontinuity at $s=D$. Using (\ref{equa:xD})
and (\ref{equa:rho_et_eta}) we find that:
\begin{equation}
\binom{\nu_e(D)}{\lambda_e(D)}  -\binom{\nu_i(D)}{\lambda_i(D)}
 = \gamma w \binom{\cos \beta}{-\sin \beta}
\label{equa:saut-de-force-au-menisque}
\end{equation}
that is the external force applied on the beam at $s=D$ is along the meniscus, as used in \citet{neukirch2007,antkowiak2011}.

Once $\gamma$, $\Delta \gamma$, $YI$, and the volume $V$ are set, the equilibrium configuration is found by solving the nonlinear boundary value problem for $s \in (0;D)$, given by Eqs.~(\ref{equa:rprime}), (\ref{equa:equilibre_moment_inside}), (\ref{equa:equilibre_moment_outside}), (\ref{equa:equilibre_force_x}), and (\ref{equa:equilibre_force_y}), with left boundary conditions (\ref{equa:BC-s0}) and right boundary conditions $\theta'(D)=0$, $\nu_i(D)=-\gamma w \cos \beta$, and $\lambda_i(D)=\gamma w \sin \beta$. The presence of unknown parameters $\beta$, $r$, and $D$ is balanced by additional conditions (\ref{equa:fixed_volume}), (\ref{equa:xD}), and (\ref{equa:YD}).

We remark that, as in the case of a rigid substrate (Section \ref{section:drop-on-rigid}), the sum of the distributed Laplace pressure $f_1=\int_0^D -\mu w x' ds$ applied along the $y$ axis on the beam (Eq. (\ref{equa:equilibre_force_y})) is balanced by the $y$ component of the meniscus force at $s=D$: $f_2=\gamma w \sin \beta$, that is $f_1+f_2=0$.

\subsection*{Equilibrium solutions}
%--------------------------------
%

\begin{figure}[ht]
    \centering
    \subfigure{\includegraphics[width=0.49\columnwidth]{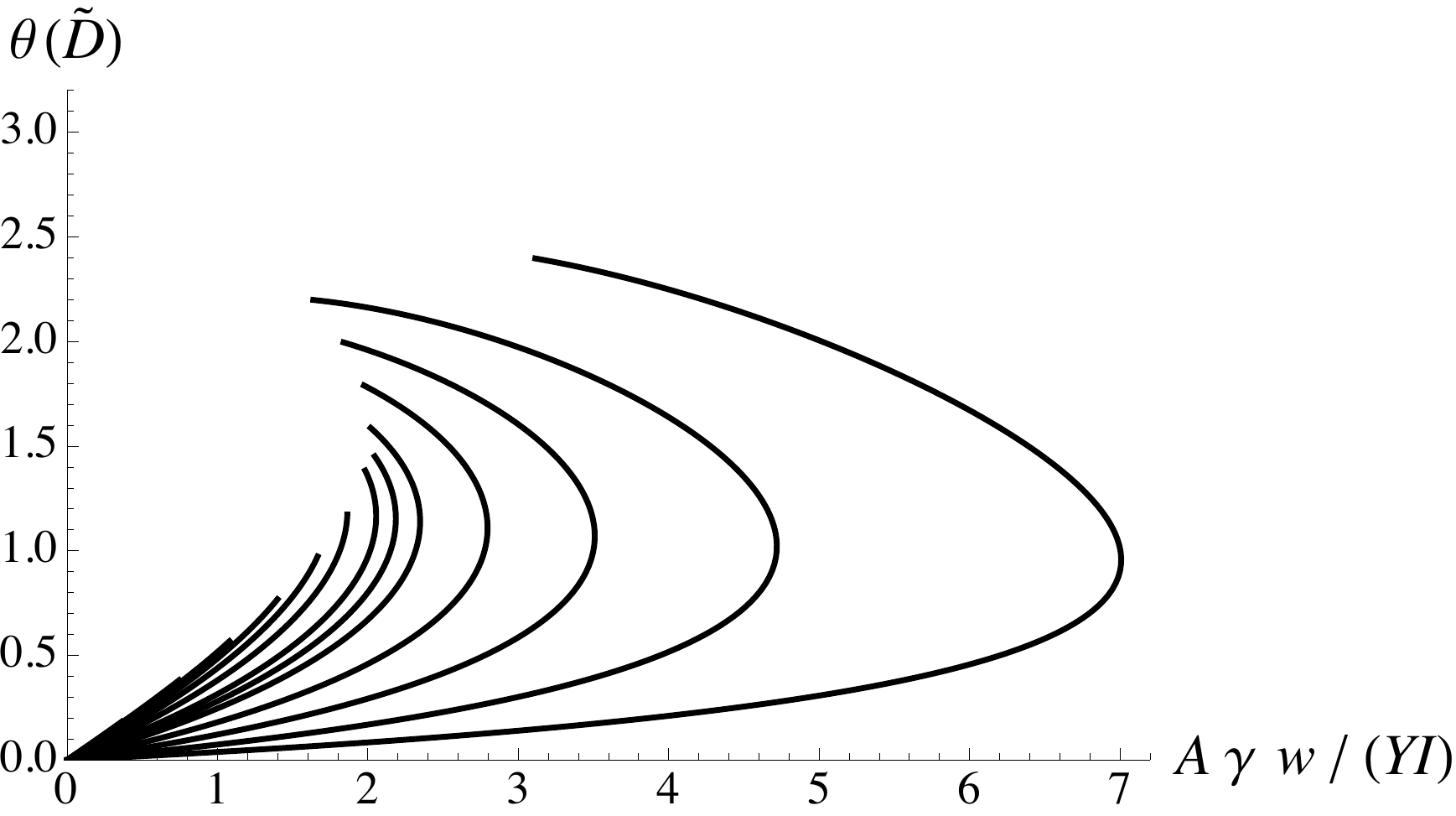}} 
    \subfigure{\includegraphics[width=0.49\columnwidth]{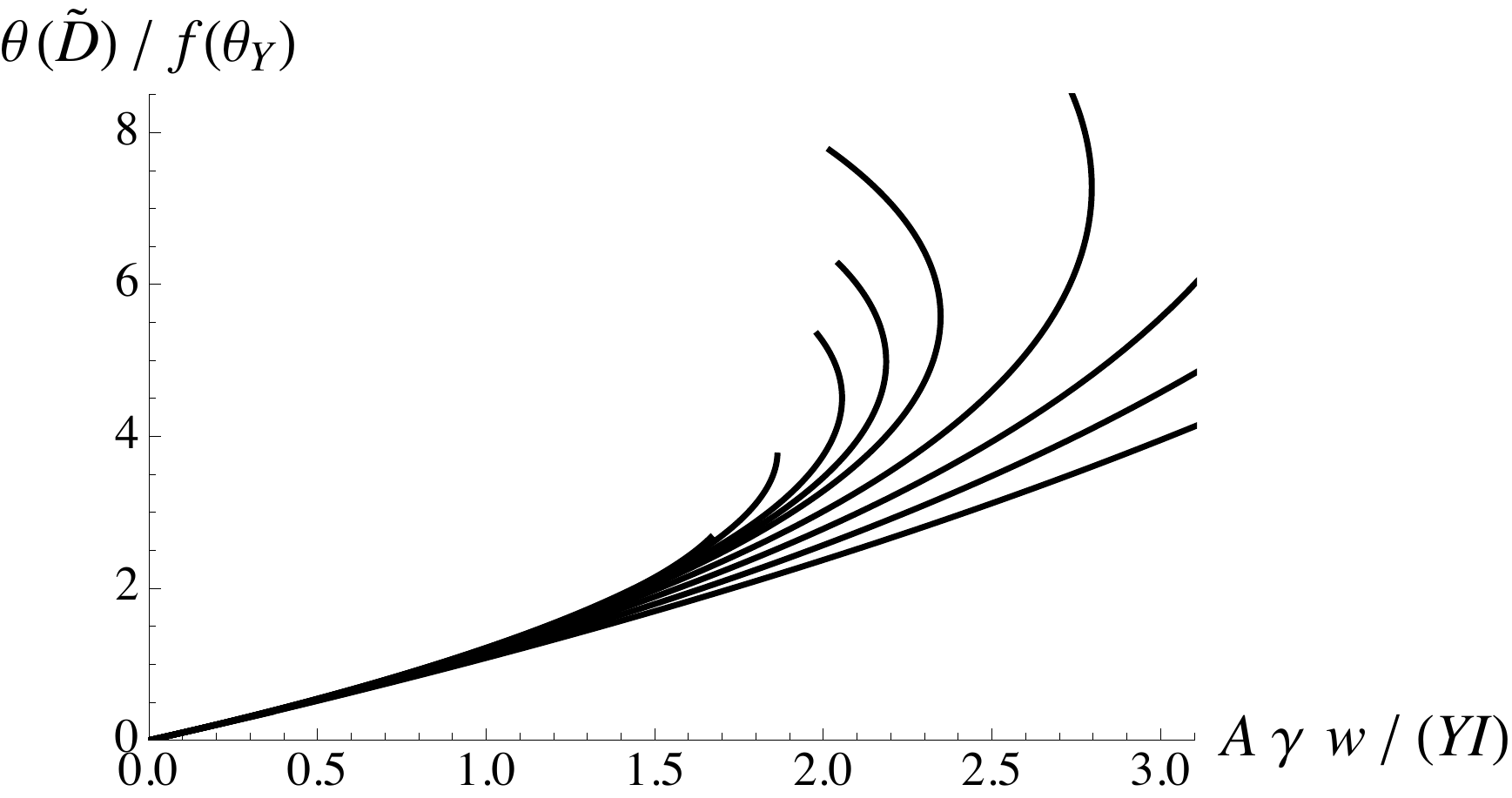}}
    \caption{Inclination $\theta(\tilde{D})$ of a beam deformed by a liquid drop, in the absence of gravity. Left: $\theta(\tilde{D})$ as a function of the non-dimensionalized surface tension $\tilde{\gamma}$, for several value of the wetting angle $\theta_Y$. Right: $\theta(\tilde{D})$ scaled with $f(\theta_Y)$, see formula (\ref{equa:thetaD-linear}). Near the origin, curves for different $\theta_Y$ collapse on a straight line of slope unity.}
    \label{fig:thetaD-no-g}
\end{figure}

We now solve the boundary value problem for different values of the parameters, e.g. $A$, $\gamma$, $\Delta \gamma$. We note $\theta_Y$ the wetting angle, defined by $\Delta \gamma + \gamma \cos \theta_Y=0$, and we use $\theta_Y$ instead of $\Delta \gamma$ as parameter. We start with non-dimensionalizing the equilibrium equations.
The configuration of the beam in the region $s \in (D;L]$ is trivial: the beam is straight and there is no stress $\theta'(s) \equiv M(s) \equiv 0$ and $\nu_e(s) \equiv \lambda_e(s) \equiv 0$. The value of the length $L$ is therefore of no importance, it can be anything as long as $L>D$. Consequently we use $\sqrt{A}$ as unit length, and $YI/A$ as unit force. The equilibrium equations for the dimensionless quantities (with over-tildes) read:
\begin{subequations}
\label{sys:equil-no-g-adim}
\begin{align}
\tilde{x}'(\tilde{s}) = \cos \theta \, 
& , \; \;  \tilde{y}'(\tilde{s}) = \sin \theta \\
\theta''(\tilde{s}) = \tilde{n}_x \sin \theta & - \tilde{n}_y \cos \theta \label{equa:equil-theta-adim}\\
\tilde{n}_x'(\tilde{s}) = - (\tilde{\gamma} / \tilde{r}) \, \sin \theta \, 
& , \; \;  \tilde{n}_y'(\tilde{s}) = (\tilde{\gamma} / \tilde{r}) \, \cos \theta
\label{equa:equil-force-adim}
\end{align}
\end{subequations}
where $\tilde{\gamma}= A \gamma w/(YI)$ is a dimensionless quantity measuring the strength of surface tension. Volume conservation (\ref{equa:fixed_volume}) now reads:
\begin{equation}
1 = \tilde{r}^2 \left(\beta - \frac{1}{2} \sin 2 \beta \right) + 2 \, \tilde{x}_D \, \tilde{y}_D - 2 \int_{0}^{\tilde{D}} \tilde{y} \, \cos \theta \, d\tilde{s} \label{equa:fixed_volume_adim}
\end{equation}
The boundary conditions at $s=0$ are $\tilde{x}(0)=0$, $\tilde{y}(0)=0$, and $\theta(0)=0$. The boundary conditions at $\tilde{s}=\tilde{D}$ are:
\begin{equation}
\tilde{x}(\tilde{D})= \tilde{r} \sin \beta \, , \;
\theta'(\tilde{D})=0 \, , \;
\tilde{n}_x(\tilde{D})=- \tilde{\gamma} \cos \beta \, , \;
\tilde{n}_y(\tilde{D})= \tilde{\gamma} \sin \beta \, , \;
\theta(\tilde{D})+\beta=\theta_Y \label{equa:BC-adim-BVP}
\end{equation}
For each value of the fixed parameters $\tilde{\gamma}$ and $\theta_Y$, we numerically solve this boundary value problem with a shooting method where the six unknowns $\theta'(0)$, $\tilde{n}_x(0)$, $\tilde{n}_y(0)$, $\beta$, $\tilde{D}$, and $\tilde{r}$ are balanced by the five boundary conditions (\ref{equa:BC-adim-BVP}) and the constraint (\ref{equa:fixed_volume_adim}).
Results for the inclination of the beam at $\tilde{s}=\tilde{D}$ are plotted in Fig.~\ref{fig:thetaD-no-g}-left. 
We remark that Eqs.~(\ref{equa:equil-theta-adim}) and (\ref{equa:equil-force-adim}) can be simplified to:
\begin{equation}
\theta'(s) = - \frac{\tilde{\gamma}}{2 r} y^2 + n_x(0) \, y -  \frac{\tilde{\gamma}}{2 r} x^2 - n_y(0) \, x + \theta'(0)
\end{equation}
and boundary conditions (\ref{equa:BC-adim-BVP}) imply $\tilde{n}_y(0)=0$.

We also look for analytical solutions when the surface tension is small $\tilde{\gamma} \ll 1$, that is when the elasto-capillary length is large: $L_{ec}=\sqrt{YI/(\gamma w)} \gg \sqrt{A}$.
We develop unknowns in power of $\tilde{\gamma}$: 
$\tilde{x}(\tilde{s}) = \tilde{x}_0(\tilde{s}) + \tilde{\gamma} \tilde{x}_1(\tilde{s}) + \ldots$ ,
$\tilde{y}(\tilde{s}) = 0 + \tilde{\gamma} \tilde{y}_1(\tilde{s}) + \ldots$ ,
$\theta(\tilde{s})=0+\tilde{\gamma} \theta_1(\tilde{s})+ \ldots$ ,
$\tilde{D}=\tilde{D}_0+\tilde{\gamma} \tilde{D}_1+ \ldots$, etc.
We find $\tilde{x}_0(\tilde{s})=\tilde{s}$, $\theta_1(\tilde{s})=\tilde{s} (3 D^2 - \tilde{s}^2) / (6 r)$ and $\tilde{y}_1(\tilde{s})=\tilde{s}^2 (6 D^2 - \tilde{s}^2) / (24 \tilde{r})$, $\beta_0=\theta_Y$, $\tilde{D}_0= \tilde{r}_0 \sin \theta_Y$, and $1/\tilde{r}_0 = \sqrt{\theta_Y - \sin \theta_Y \, \cos \theta_Y}$. This yields:
\begin{align}
\theta(\tilde{D}) =& \frac{\tilde{\gamma}}{3} \, \frac{\sin^3 \theta_Y}{\theta_Y - \sin \theta_Y \, \cos \theta_Y} + O(\tilde{\gamma}^2)  = \tilde{\gamma} \, f(\theta_Y) + O(\tilde{\gamma}^2) \label{equa:thetaD-linear} \\
\tilde{y}(\tilde{D}) =& \frac{5\tilde{\gamma}}{24} \, \frac{\sin^4 \theta_Y}{\left(\theta_Y - \sin \theta_Y \, \cos \theta_Y\right)^{3/2}} + O(\tilde{\gamma}^2) 
\end{align}

%
%===========================================%
%
%
\section{Liquid drop on a flexible beam, in the presence of gravity} \label{section:with-gravity}
%==========================
%
%
%
\begin{figure}[ht]
    \centering
    \includegraphics[width=.7\columnwidth]{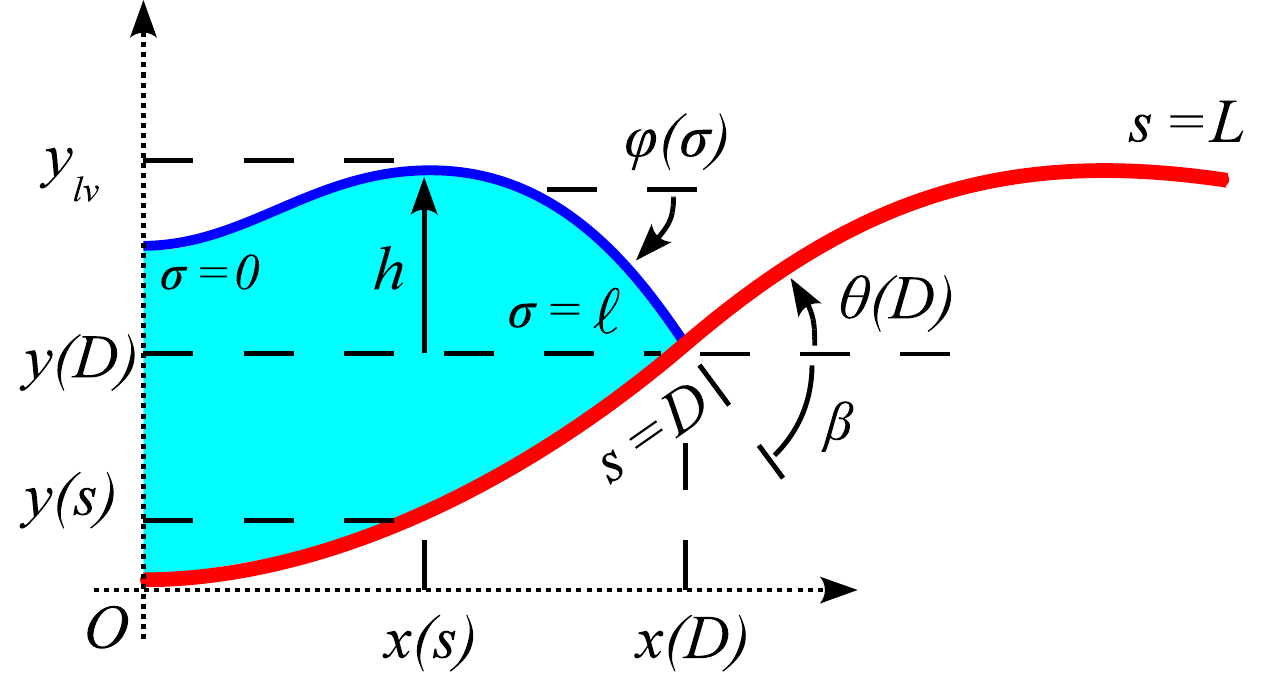}
    \caption{A flexible beam bent by a liquid drop, in the presence of gravity. Due to symmetry, we only consider positive $x$'s.}
    \label{fig:config-with-g}
\end{figure}
We now consider the situation where gravity is no longer neglected. The beam has mass per length $\tau$ and the weight of the beam introduces the term:
\begin{equation}
\int_0^L \tau g y(s) ds
\end{equation}
in the potential energy (\ref{equa:lagrangian2}).
Yet the main difference is that the liquid-air interface is no longer a circular-arc, and its shape $y_{lv}(x)$ has to be solved for. The liquid-air interface energy is now:
\begin{equation}
\gamma w \int_0^{x(D)} \sqrt{1+y_{lv}'(x)^2} \; \d x + 2 \gamma  A
\end{equation}
In addition the weight of the liquid (of density $\rho$) has to be accounted for, yielding the term:
\begin{equation}
\rho g w \int_0^{x(D)} \int_y^{y_{lv}} y \, \d y \, \d x
\end{equation}
Finally, the constraint of constant volume now reads:
\begin{equation}
A = \int_0^{x(D)} \int_y^{y_{lv}}  \d y \, \d x
\end{equation}
As done previously, the beam geometry is parametrized with the arc-length $s$: $x(s)$, $y(s)$, $\theta(s)$.
As for the liquid-air interface, we introduce the relative height $h = y_{lv} - y(D)$ and  we parametrize it with the same variable $s$: $h(s)=h(x(s))$, see Fig.~\ref{fig:config-with-g}.
The arc-length $s$ along the beam is thus the unique independent variable.
Dropping out constant terms in the energy, the Lagrangian of the system reads:
\begin{align}
{\cal L}(x,y,\theta,h,D) &=
\frac{1}{2} YI \int_{0}^{L} \left[ \theta'(s) \right]^2 \d s   
+ \int_{0}^{L} \tau g y  \d s  +  w D \Delta \gamma
+ \gamma w \int_0^D \sqrt{x'^2+h'^2}\, \d s \nonumber  \\
& + \rho g w \int_0^D \frac{1}{2} \left[ \left(h+y(D)\right)^2 - y^2 \right] x' \d s
- \mu w \int_0^D (h+y(D) - y) x' \d s  \nonumber \\
&  + \int_{0}^{D} \nu_i(s) \,  \left[x'-\cos \theta \right] \d s + \int_{D}^{L} \nu_e(s) \,  \left[x'-\cos \theta \right] \d s \nonumber \\
& + \int_{0}^{D} \lambda_i(s) \,  \left[y'-\sin \theta \right] \d s+ \int_{D}^{L} \lambda_e(s) \,  \left[y'-\sin \theta \right] \d s 
\label{equa:lagrangian-with-g}
\end{align}
Boundary conditions are $x(0)=y(0)=\theta(0)=0$, $h'(0)=0$, and $h(D)=0$. 
As in Section \ref{section:flex-no-g}, we note $X=(x,y,\theta,h,D)$ and we look for $X_e$ such that the first variation ${\cal L}'(X_e)(\bar{X})$ vanishes when $X$ is changed from $X_e$ to $X_e+\epsilon \bar{X}$.
We compute the first variation
of ${\cal L}$ with regard to the variable $X=(x,y,\theta,h,D)$:
%=X_e+\epsilon \bar{X}$:
%
\begin{align}
{\cal L}'(X_e)&(\bar{X}) 
= YI \int_{0}^{L} \theta' \, \bar{\theta}' \d s   
+ \int_{0}^{L} \tau g \bar{y} \d s   
+  w \bar{D} \Delta \gamma \nonumber \\
& + \gamma w \int_0^D \frac{x' \bar{x}' + h' \bar{h}'}{\sqrt{x'^2+h'^2}} \, \d s 
+ \bar{D} \gamma w \sqrt{x'(D)^2+h'(D)^2}\nonumber \\
& + \rho g w \int_0^D \Biggl\{ h \bar{h} x' + h \bar{y}(D) x' + h y'(D) \bar{D} x'  + \bar{h} y(D) x' + y(D) \bar{y}(D) x' + \nonumber \\
& \qquad 
\bar{D} y(D) y'(D) x' - y \bar{y} x' +  \frac{1}{2} \left[ \left(h+y(D)\right)^2 - y^2 \right] \bar{x}' \Biggr\} \, \d s \nonumber \\
& - \mu w \int_0^D \left[ \left(\bar{h}+\bar{y}(D)+\bar{D} y'(D) - \bar{y} \right) x' + \left(h+y(D) - y \right) \bar{x}' \right] \d s \nonumber \\
&  + \int_{0}^{D} \nu_i(s) \,  \left[\bar{x}'+ \bar{\theta}\sin \theta \right] \d s + \int_{D}^{L} \nu_e(s) \,  \left[\bar{x}'+ \bar{\theta}\sin \theta \right] \d s \nonumber \\
& + \int_{0}^{D} \lambda_i(s) \,  \left[\bar{y}'-\bar{\theta}\cos \theta \right] \d s+ \int_{D}^{L} \lambda_e(s) \,  \left[\bar{y}'-\bar{\theta}\cos \theta \right] \d s 
\label{equa:1st-variation-expression-with-g}
\end{align}
We now require the first variation to vanish for all $\bar{x},\bar{y},\bar{\theta},\bar{h}$, and $\bar{D}$.
Collecting terms involving $\bar{h}$ and $\bar{h}'$ yields, after integration by parts:
\begin{equation}
\gamma w \left[ \frac{h' \bar{h}}{\sqrt{x'^2+h'^2}}\right]_0^D + \int_0^D \left( -\gamma w \left( \frac{h'}{\sqrt{x'^2+h'^2}}\right)' + \rho g w (h+y(D))x' - \mu w x'\right) \bar{h} \, \d s \label{equa:h-bar}
\end{equation}
The first term is $\gamma w h'(D) \bar{h}(D) / \sqrt{x'(D)^2+h'(D)^2}$. Boundary conditions require that $h+\epsilon \bar{h}$ vanishes at $s=D+\epsilon \bar{D}$; this yields $\bar{h}(D)= - \bar{D} h'(D)$. Consequently this first term effectively goes into the equation for $\bar{D}$, see (\ref{equa:1st-varia-for-all-D}).
The second term of (\ref{equa:h-bar}) has hence to vanish for all $\bar{h}(s)$, which implies that the liquid-air interface $h(s)$ obeys the equation:
\begin{equation}
\left[ \rho g w (h+y(D)) - \mu w \right] x' = \gamma w \left( \frac{h'}{\sqrt{x'^2+h'^2}}\right)' \label{equa:interface}
\end{equation}
Integrating this equation from $s=0$ to $s=D$ yields:
\begin{equation}
\rho g w \hat{A} - \mu w x(D) = \gamma w \sin \varphi(D) \label{equa:hat-A} 
\end{equation}
where $\hat{A}=\int_0^D (h+y(D)) x' \d s$ is the area between the liquid-air interface and the horizontal axis, and where $\varphi$ is the angle the interface makes with the horizontal.
Evaluating (\ref{equa:interface}) at $s=0$ reveals that the Lagrange multiplier $\mu$ is the hydrostatic pressure at the origin.

Requiring (\ref{equa:1st-variation-expression-with-g}) to vanish for all $\bar{D}$ yields:
\begin{equation}
w \Delta \gamma + w \gamma \sqrt{x'(D)^2+h'(D)^2} 
- w \gamma \frac{h'(D)^2}{\sqrt{x'(D)^2+h'(D)^2}} +y'(D) \left(\rho g w  \hat{A} - \mu w x(D)\right)=0
\label{equa:1st-varia-for-all-D}
\end{equation}
Using $x'(D)/\sqrt{x'(D)^2+h'(D)^2} = \cos \varphi(D)$, $x'(D)=\cos \theta(D)$, $y'(D)=\sin \theta(D)$, and (\ref{equa:hat-A}) we arrive at:
\begin{equation}
\Delta \gamma + \gamma \cos \left[ \theta(D) + \beta \right] = 0
\end{equation}
where $\beta = - \varphi(D)$. This is Young-Dupr\'e relation for the wetting angle $\theta(D) + \beta$.
We collect terms involving $\bar{x}$ and $\bar{x}'$ in (\ref{equa:1st-variation-expression-with-g}) and we integrate by parts to obtain:
\begin{align}
& \left[ \left\{ \frac{w \gamma x' }{\sqrt{x'^2+h'^2}} + \frac{1}{2} \rho g w \left( h+y(D)\right)^2 - \frac{1}{2} \rho g w y^2+ \nu_i -\mu w (h+y(D)-y)  \right\} \bar{x} \right]_0^D \nonumber \\
&-\int_0^D \left\{  \left(  \frac{w \gamma x' }{\sqrt{x'^2+h'^2}} \right)'  + \rho g w (h+y(D)) h' -\rho g w y y' - \mu w (h'-y') + \nu'_i  \right\} \bar{x} \, \d s \nonumber \\
& + \Bigl[ \nu_e \, \bar{x} \, \Bigr]_D^L -\int_D^L  \nu_e' \, \bar{x} \, \d s =0
\end{align}
The fact that we have $\bar{x}(0)=0$, but arbitrary $\bar{x}(D)$ and $\bar{x}(L)$ implies:
\begin{align}
\nu_e(L) &= 0 \\
\nu'_e(s) &= 0 \\
 \nu_e(D) &- \nu_i(D) = w \gamma x'(D)/\sqrt{x'(D)^2+h'(D)^2} 
 = w \gamma \cos \varphi(D)= w \gamma \cos \beta \label{equa:saut_force_x-with-g} \\
 \nu'_i &= - \left(\mu w - \rho g w y  \right) y' - \left[ \rho g w (h+y(D)) - \mu w \right] h' - \gamma w \left( \frac{x'}{\sqrt{x'^2+h'^2}}\right)' 
 \label{equa:equilibre_force_x-with-g}
\end{align}
Considering the identity $ x' \left( x' / \sqrt{x'^2+h'^2} \right)' + h' \left( h' / \sqrt{x'^2+h'^2} \right)' = 0 $ and (\ref{equa:interface}), equation (\ref{equa:equilibre_force_x-with-g}) reduces to:
\begin{equation}
\nu'_i = - \left(\mu w - \rho g w y  \right) y'   \label{equa:equilibre_force_x-with-g-bis}
\end{equation}
Requiring (\ref{equa:1st-variation-expression-with-g}) to vanish for all $\bar{y}$ similarly yields, after the use of (\ref{equa:hat-A}):
\begin{align}
\lambda_e(L) = 0 \, &, \; \;  
\lambda_e(D) - \lambda_i(D) = - w \gamma \sin \beta \label{equa:saut_force_y-with-g} \\
\lambda'_e = \tau g \, &, \; \;  \lambda'_i =  \tau g + (\mu w-\rho w g y) x' 
\label{equa:equilibre_force_y-with-g}
\end{align}
Finally, requiring (\ref{equa:1st-variation-expression-with-g}) to vanish for all $\bar{\theta}$ yields the same equation as before, see Eqs. (\ref{equa:equilibre_moment_inside}) and (\ref{equa:equilibre_moment_outside}).

From (\ref{equa:saut_force_x-with-g}) and (\ref{equa:saut_force_y-with-g}), we see that the internal force experience the same discontinuity as in the case without gravity (\ref{equa:saut-de-force-au-menisque}): here also the external force applied on the beam at $s=D$ is along the meniscus.

\subsection*{Equilibrium solutions}
%--------------------------------
%

We now solve the boundary value problem for different values of the parameters, e.g. $A$, $\gamma$, $\Delta \gamma$, $\tau$. We note $\theta_Y$ the wetting angle, defined by $\Delta \gamma + \gamma \cos \theta_Y=0$, and we use $\theta_Y$ instead of $\Delta \gamma$ as parameter. We start with non-dimensionalizing the equilibrium equations.
As the configuration of the beam in the region $s \in (D;L]$ is no longer trivial, we use $L$ as unit length, $EI/L^2$ as unit force, and $EI/L$ as unit moment. For the beam, the equilibrium equations for the dimensionless quantities (with over-tildes) read:
\begin{subequations}
\label{sys:equil-with-g-adim}
\begin{align}
\tilde{x}'(\tilde{s}) = \cos \theta \, 
& , \; \;  \tilde{y}'(\tilde{s}) = \sin \theta \\
\theta''(\tilde{s}) = \tilde{n}_x \sin \theta & - \tilde{n}_y \cos \theta \\
\tilde{n}_x'(\tilde{s}) = - P \, \sin \theta & , \; \;  
\tilde{n}_y'(\tilde{s}) = \tilde{\tau} + P \, \cos \theta
\end{align}
\end{subequations}
where $P$ is the dimensionless hydrostatic pressure $P= (L/L_{ec})^2 \, [ \tilde{\mu} - (L/L_c)^2 \, \tilde{y}]$ for the region $\tilde{s} \in [0;\tilde{D})$ and $P=0$ for $\tilde{s}>\tilde{D}$.  We have introduced the dimensionless pressure $\tilde{\mu}=\mu L/\gamma$, the capillary length $L_c= \sqrt{\gamma/(\rho g)}$, and the elasto-capillary length $L_{ec}=\sqrt{YI/(\gamma w)}$ \citep{bico+al:2004}.

The equations for the liquid-air interface (\ref{equa:interface}) can be rewritten using $(i)$ the angle $\varphi$ the interface does with the horizontal, and $(ii)$ the arc-length $\sigma$ along this interface:
\begin{subequations}
\label{sys:equil-with-g-adim-interface}
\begin{align}
\varphi'(\tilde{\sigma}) &= (L/L_c)^2 \, \left[ \tilde{h}+\tilde{y}(\tilde{D}) \right] - \tilde{\mu} \\
 \tilde{h}'(\tilde{\sigma}) &= \sin \varphi  \; , \: \:
 \tilde{\xi}'(\tilde{\sigma}) = \cos \varphi 
\end{align}
\end{subequations}
where $ \tilde{\sigma}=\sigma/L$. The liquid-air interface has total contour length $\ell$.

As soon as values for the fixed parameter $L_{ec}/L$, $L_c/L$, $\tilde{\tau}$, $\tilde{A}$, and $\theta_Y$ are given, the boundary value problem can be solved with a shooting procedure where $\theta'(0)$, $\tilde{n}_x(0)$, $\tilde{n}_y(0)$, $\tilde{\mu}$, $\tilde{h}(0)$, $\tilde{D}$, $\tilde{\ell}$ are seven unknowns.
Integration of (\ref{sys:equil-with-g-adim}) is performed for $\tilde{s} \in [0;\tilde{D})$ and integration of (\ref{sys:equil-with-g-adim-interface}) is performed for $\tilde{\sigma} \in [0;\tilde{\ell}]$. At $\tilde{s}=\tilde{D}$ a jump in the force vector is introduced according to (\ref{equa:saut_force_x-with-g}) and (\ref{equa:saut_force_y-with-g}) with $\beta=-\varphi(\tilde{\ell})$ and $n_x=\nu$, $n_y=\lambda$. Then integration of (\ref{sys:equil-with-g-adim}) is performed for $\tilde{s} \in (\tilde{D};1)$.
Seven boundary equations have then to be fulfilled: 
%$\theta'(1)=0$, $\tilde{n}_x(1)=0$, $\tilde{n}_y(1)=0$,  $\tilde{h}(\tilde{\ell})=0$, $\tilde{\xi}(\tilde{\ell})=\tilde{x}(\tilde{D})$, $\theta(\tilde{D})+\beta=\theta_Y$, and the volume condition $\tilde{A}=\int_0^{\tilde{\ell}} \tilde{h} \cos \varphi \, d\tilde{\sigma} - \int_0^{\tilde{D}} (\tilde{y}-\tilde{y}(\tilde{D})) \cos \theta \, d\tilde{s}$.
\begin{equation}
\theta'(1)=0 \, , \: \tilde{n}_x(1)=0 \, , \: \tilde{n}_y(1)=0 \, , \: \tilde{h}(\tilde{\ell})=0 \, , \: \tilde{\xi}(\tilde{\ell})=\tilde{x}(\tilde{D}) \, , \:  \theta(\tilde{D})+\beta=\theta_Y
\label{equa:BC-system-complet}
\end{equation}
together with the volume condition:
\begin{equation}
\tilde{A}=\int_0^{\tilde{\ell}} \tilde{h} \cos \varphi \, d\tilde{\sigma} - \int_0^{\tilde{D}} \left[ \tilde{y}-\tilde{y}(\tilde{D}) \right] \cos \theta \, \d \tilde{s}
\label{equa:volume-system-complet}
\end{equation}

\begin{figure}[ht]
    \centering
    \includegraphics[width=0.7\columnwidth]{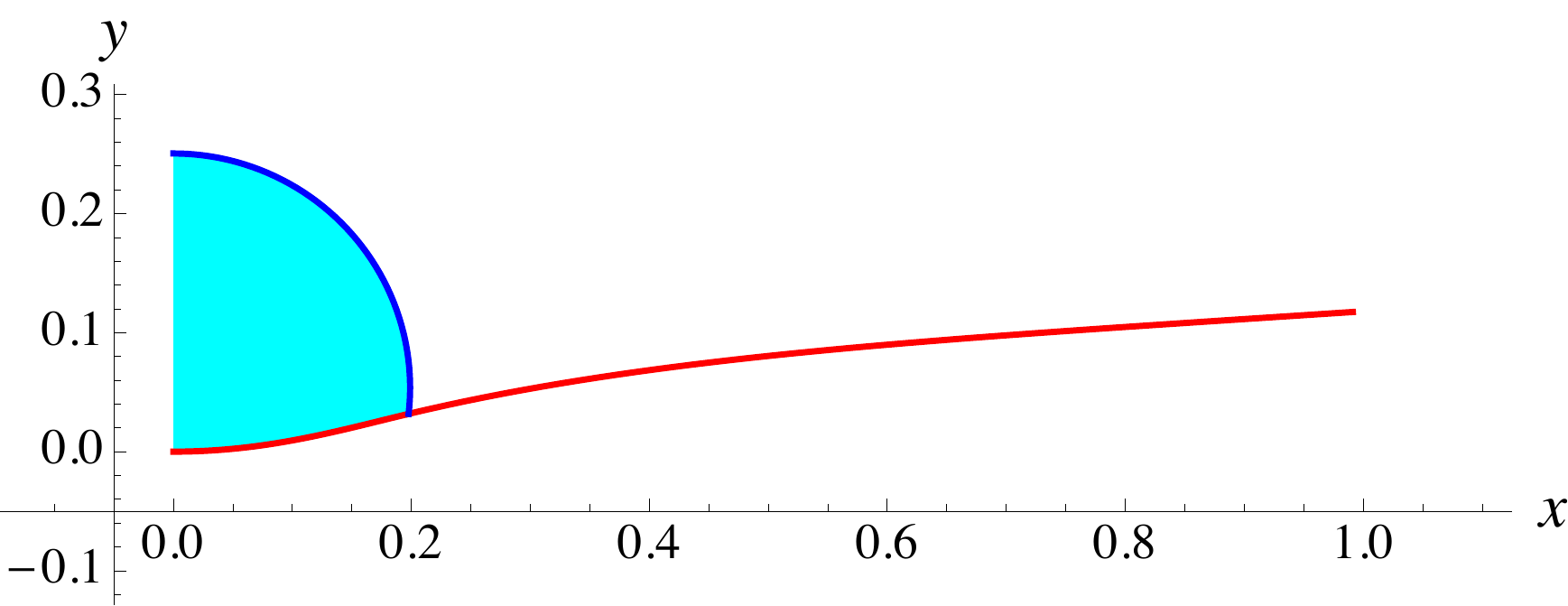}
    \caption{A flexible beam bent by a liquid drop, in the presence of gravity.}
    \label{fig:solution-with-g}
\end{figure}

A solution for $L_{ec}/L=0.175$, $L_c/L=0.982$, $\tilde{\tau}=2$, $\tilde{A}=0.039$, and $\theta_Y=110$ deg. is shown in Fig.~\ref{fig:solution-with-g}.
The seven unknowns of the shooting procedure are found to be $\theta'(0)=1.887$, $\tilde{n}_x(0)=9.15$, $\tilde{n}_y(0)=-3.32$, $\tilde{\mu}=5.24$, $\tilde{h}(0)=0.219$, $\tilde{D}=0.201$, and $\tilde{\ell}=0.333$.
%
%===========================================%
%
%
\section{Pinning, receding, and advancing of the contact line} \label{section:contact-line-pinning}
%==========================
%
%
%
We now turn to the case of contact line pinning and we show, following \citet{Alberti-DeSimone:Quasistatic-Evolution-of-Sessile:2011}, that pinning, receding, and advancing of the contact line can be treated in a variational approach where irreversibility conditions are introduced.
We consider the drop of Fig.~\ref{fig:young-dupre}, lying at equilibrium on a rigid substrate with its  contact angle $\beta$ equal to the Young-Dupr\'e angle $\theta_Y$, defined by $\Delta \gamma + \gamma \cos \theta_Y=0$, $\beta(t=0) = \theta_Y$.
At time $t=0$ evaporation starts to take place and we study the subsequent behavior of the drop.
In the absence of contact line pinning, the contact angle $\beta$ will stay at $\beta(t)=\theta_Y$, and the wetted length $D$ will decrease in order to keep on fulfilling the volume constraint.
In the presence of contact line pinning, the length $D$ will first stay fixed and the contact angle $\beta$ will decrease: $\beta(t>0) < \theta_Y$ to  fulfill the volume constraint. Eventually, as $\beta(t)$ reaches a receding threshold, the contact line will start to move: $D=D(t)$.

As in section \ref{section:drop-on-rigid} we start with the interfaces energies, Eq.~\ref{equa:interface-energy}, and we utilize the constraint $D=r \, \sin \beta$ to eliminate the variable $r$: per unit $w$ the energy is $2 \gamma D \beta / \sin \beta + 2 D \Delta \gamma$.
%\begin{equation}
%2 \gamma D \frac{\beta}{\sin \beta} + 2 D \Delta \gamma
%\end{equation}
As $D$ decreases from $D(t=0)$ receding of the contact line is associated with an energy dissipation $k>0$ per unit area \citep{Joanny-Gennes:A-model-for-contact-angle:1984,Alberti-DeSimone:Wetting-of-rough-surfaces::2005}, which we introduce in the energy:
\begin{equation}
E(D,\beta)= 2 \gamma D \frac{\beta}{\sin \beta} + 2 D \Delta \gamma + k (D(0)-D)
\end{equation}
We minimize this energy under $(i)$ the constraint of fixed volume $V/w = r^2 (\beta - \sin \beta \, \cos \beta)=(D/\sin \beta)^2 (\beta - \sin \beta \, \cos \beta)$, and $(ii)$ the irreversibility condition $D(t^+) \leq D(t^-)$, which we note $D \leq D^-$.
We therefore introduce the Lagrangian:
\begin{equation}
{\cal L}=E(D,\beta) - \lambda (D^- - D) - \mu D^2 \left( \frac{\beta}{\sin^2 \beta} - \frac{\cos \beta}{\sin \beta} \right)
\end{equation}
where $\mu$ is the Lagrange multiplier associated with the volume equality constraint and where $\lambda$ is the Lagrange multiplier associated with the inequality constraint $D \leq D ^-$.
The necessary conditions for having a minimum are:
\begin{align}
\frac{\partial {\cal L}}{\partial \beta}=0 &\: \Rightarrow \: \mu D = \gamma \sin \beta \\
\frac{\partial {\cal L}}{\partial D}=0 &\: \Rightarrow \: 2 \gamma \cos \beta + 2 \Delta \gamma -k +\lambda=0 \label{equa:pinning-eq-b}\\
\mbox{K.T.} & \: \Rightarrow \: \lambda \geq 0 \, , \: (D^- -D) \geq 0 \, , \, \mbox{and } \lambda (D^- -D) = 0
 \label{equa:pinning-eq-c}
\end{align}
where the last line lists the classical Kuhn-Tucker conditions \citep{Luenberger:Introduction-to-linear-and-nonlinear:1973} arising in case of inequality constraints.
These three conditions express the fact that either the pinning force $\lambda$ is zero  and sliding occurs $D<D^-$, or the pinning force is strictly positive $\lambda>0$ preventing the contact line to move: $D=D^-$. We now introduce an angle $\theta^\star$ such that $2 (\Delta \gamma + \gamma \cos \theta^\star) = k$. Positivity of the dissipation $k$ implies that $\theta^\star < \theta_Y$. Equation (\ref{equa:pinning-eq-b}) becomes $2 \gamma (\cos \beta - \cos \theta^\star) +\lambda = 0$ and eliminating $\lambda$ in (\ref{equa:pinning-eq-c}), we finally obtain:
\begin{subequations}
\begin{align}
\cos \theta^\star - \cos \beta &\geq 0 \\
D^- - D &\geq 0\\
(\cos \theta^\star - \cos \beta) (D^- - D) & = 0
\end{align}
\end{subequations}
which means that either $D$ is fixed to $D^-$ and $\beta > \theta^\star$ (contact line pinning) or $\beta = \theta^\star$ and $D$ decreases $D < D^-$ (contact line sliding, here receding), as was used in \cite{Rivetti-Neukirch:Instabilities-in-a-drop-strip-system::2012}.

Note that for simplicity we have only presented equations for receding of the contact line, but the present treatment can be done for the general case where advancing and receding can both occur, see \citet{Fedeli-Turco:Metastable-equilibria-of-capillary:2011}.

%
%===========================================%
%
%
\subsection*{Illustration} \label{section:illustration}
\begin{figure}[ht]
\centering
\includegraphics[width=14.5cm]{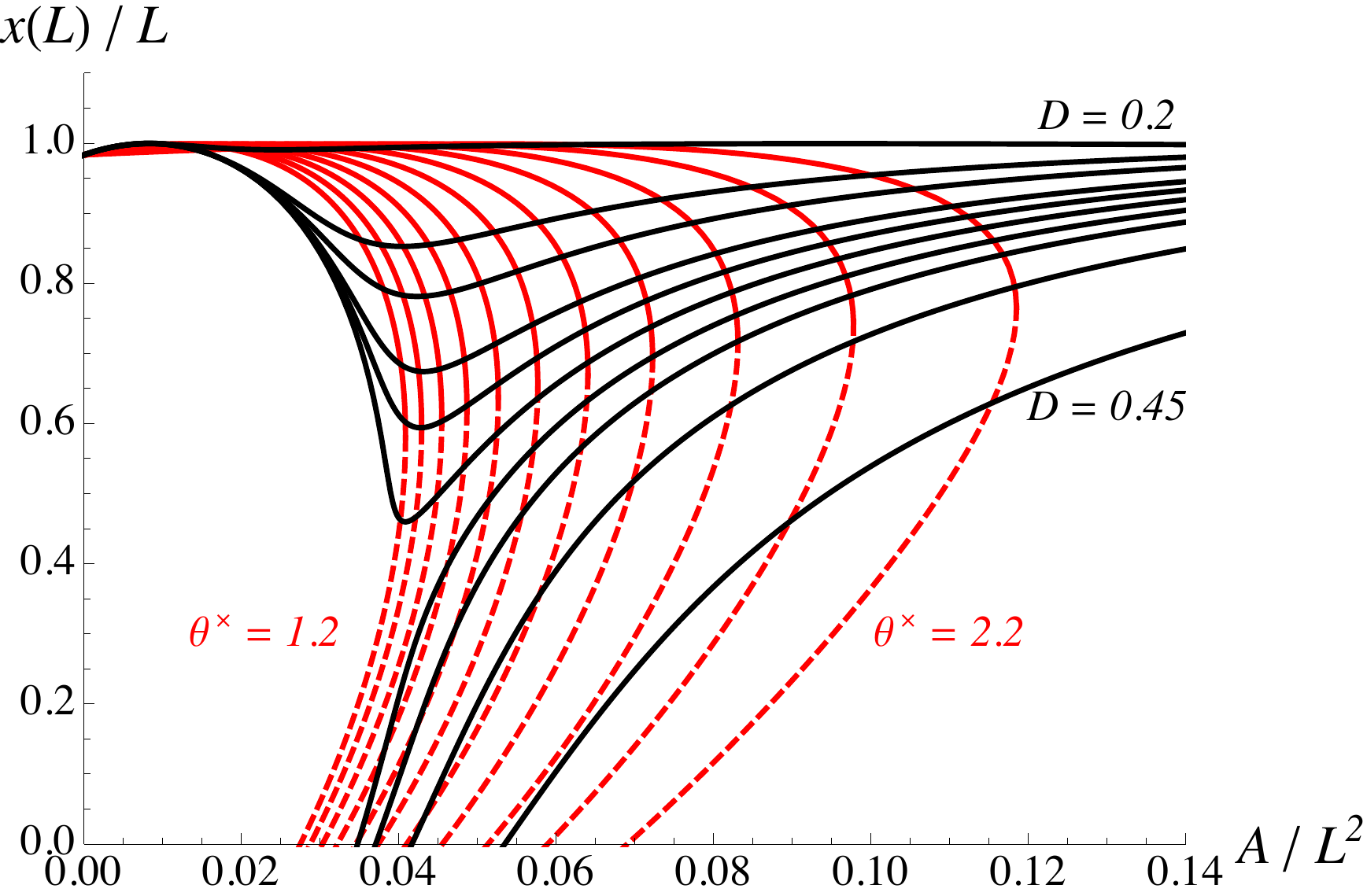}
\caption{Bifurcation diagram for the system of Fig.~\ref{fig:config-with-g} with $L_{ec}=0.2 \, L$, $L_c=0.8 \, L$, and $\tilde{\tau}=1.4$. The first set of eleven (red) curves correspond to equilibrium in the sliding hypothesis $\theta(D)+\beta=\theta^\star$, with $\theta^\star=1.2$, 1.3, $\ldots$, 2.2. These curves all have a limit point for the variable $\tilde{A}=A/L^2$. Dashed parts of the curves correspond to unstable equilibria.
The second set of ten (black) curves correspond to equilibrium in the pinning hypothesis $D=D(0)$, with $D(0)/L=0.2$, 0.3, 0.32, 0.34, 0.35, 0.36, 0.37, 0.38, 0.4, 0.45.}
\label{fig:bif-diag-complet}
\end{figure}

%==========================
%
%
%
We here illustrate the present theory on an imaginary experiment where one deposits a drop on an elastic strip (see Fig.~\ref{fig:config-with-g}) and wait for evaporation to take place \citep{Py-Reverdy-Capillary-Origami:-Spontaneous-2007}. After deposition the drop contact angle $\theta(D)+\beta$ takes some intermediate value between receding ($\theta^\star$) and advancing values.
We fix parameters $L_{ec}=0.2 \, L$, $L_c=0.8 \, L$, and $\tilde{\tau}=1.4$ and we first solve equations (\ref{sys:equil-with-g-adim}), (\ref{sys:equil-with-g-adim-interface}), (\ref{equa:BC-system-complet}), and (\ref{equa:volume-system-complet}) for several values of the receding angle $\theta^\star$ in the sliding hypothesis $\theta(D)+\beta=\theta^\star$. We then solve the equations for several values of $D(0)$ in the pinning hypothesis $D=D(0)$. Results are shown in Fig.~\ref{fig:bif-diag-complet}.

\begin{figure}[ht]
    \centering
    \includegraphics[width=15cm]{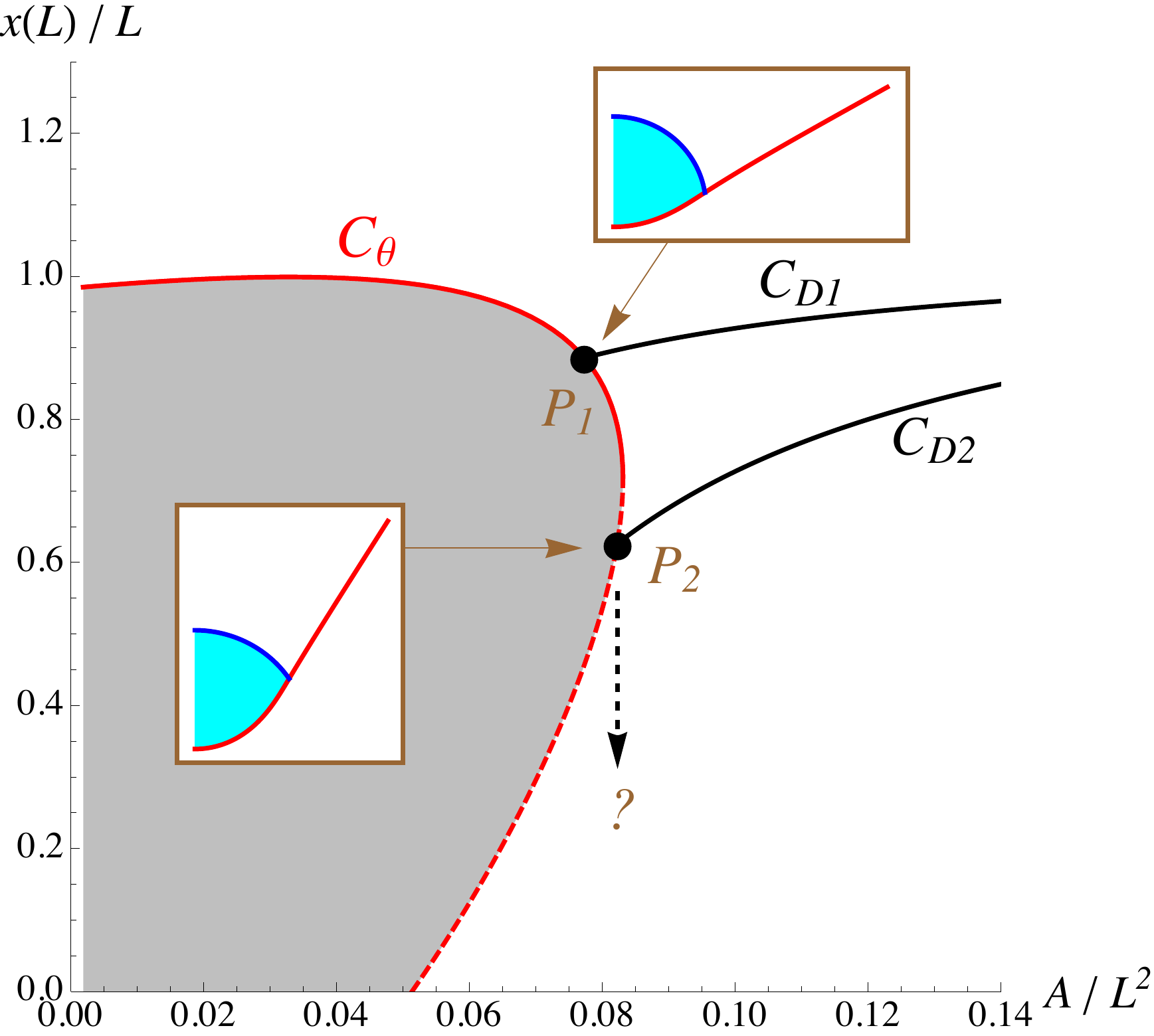}
    \caption{Evaporation experiment with $L_{ec}=0.2 \, L$, $L_c=0.8 \, L$, and $\tilde{\tau}=1.4$. The curve $C_\theta$ corresponds to equilibrium in the sliding hypothesis $\theta(D)+\beta=\theta^\star$ with $\theta^\star=2$; the region $\theta(D)+\beta < 2$ is shown shaded.
The curves $C_{D1}$ and $C_{D2}$ correspond to equilibrium in the pinning hypothesis $D=D(0)$, with $D(0)/L=0.32$, and 0.4 respectively. If evaporation starts on $C_{D1}$ at $A/L^2=0.14$ the system reaches point $P_1$ and then follows the upper part of curve $C_\theta$ down to $A=0$. If evaporation starts on $C_{D2}$ at $A/L^2=0.14$ the system reaches point $P_2$. As configurations on $C_\theta$ are unstable in this region, the system jumps to a configuration with the same value of $A$, to be discussed elsewhere.}
    \label{fig:typical-experiment}
\end{figure}

In a typical experiment, starting at $\tilde{A}=A/L^2=0.14$ with $D(0)/L=0.32$, evaporation first results in the decrease of the contact angle, following the curve $C_{D1}$ in Fig.~\ref{fig:typical-experiment}. As the contact angle $\theta(D)+\beta$ reaches $\theta^\star$ (with say $\theta^\star=2$) the system switches branch at point $P_1$ and follows the constant contact angle curve $C_\theta$ down to $\tilde{A}=0$. If now one starts at $\tilde{A}=0.14$ with $D(0)/L=0.4$, during evaporation the system follows curve $C_{D2}$ down to point $P_2$. The constant contact angle curve at $P_2$ being unstable the system jumps to a configuration with same volume, not discussed here.

%

%
%===========================================%
%
%
\section{Conclusion} \label{section:conclusion}
%==========================
%
%
%
In conclusion we have recalled that the classical Young-Dupr\'e relation for the contact angle of a drop lying on a rigid substrate can be derived from a variational approach where the concept of force is not invoked.
The variational approach has then been extended to the case where the substrate is a flexible beam and we have shown that $(i)$ Young-Dupr\'e relation still holds, and that $(ii)$ the external force applied on the elastic beam at the triple point is tangential to the liquid-vapor interface.
We then extended the approach to the case where gravity is included and found that these two results continue to hold.
Finally we have illustrated our model with the study of the evaporation of a drop deposited on a flexible beam and we have shown that, depending on the initial spreading of the drop on the beam, evaporation leads to a flat or a folded system.

The present result showing that in the case a flexural deformations the external force on the elastic beam is along the meniscus is in contradiction to what is found in \citet{Marchand-Das:Capillary-Pressure-and-Contact:2012} in the case of extensional deformation. We are now working on the extension of the present variational approach to the case of extensional deformations and hope to resolve this apparent disagreement.

%\subsection*{Acknowledgments}
\ack{
This work was supported by ANR grant  ANR-09-JCJC-0022-01.
Financial support from `La Ville de Paris - Programme \'Emergence' is also gratefully acknowledged.}
%\end{acknowledgements}

\appendix

\section{Appendix 1: Planar elastica} \label{appendix:elastica}

\begin{figure}[ht]
    \centering
    \includegraphics[width=8cm]{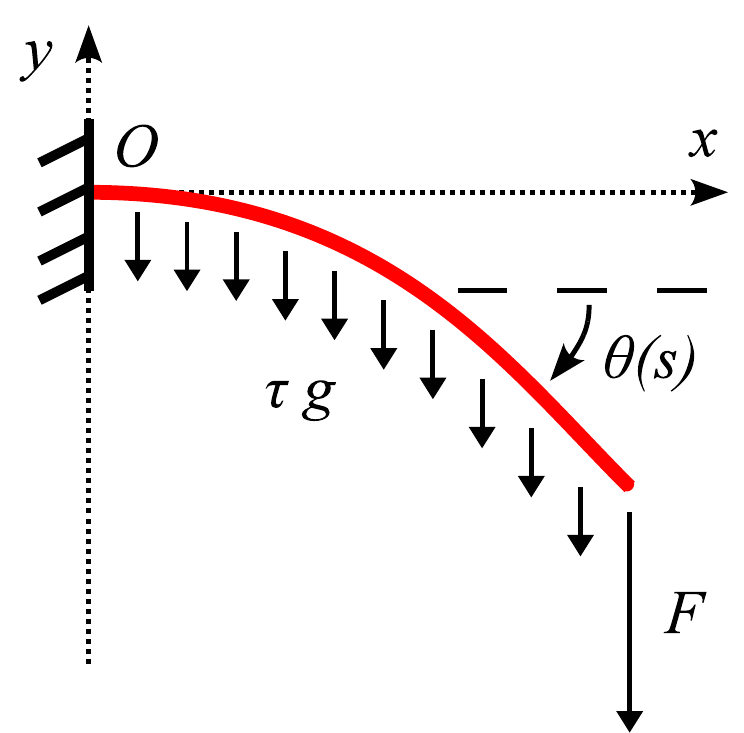}
    \caption{Cantilever beam sagging under the combined actions of its own weight $\tau \, g$ and of a localized shear force $F$ at its left end.}
    \label{fig:planar-elastica}
\end{figure}
The equilibrium equations for the system in Fig.~\ref{fig:planar-elastica} are \citep{love:1944,antman2004}:
\begin{subequations}
\label{sys:planar-elastica}
\begin{align}
x'(s) = \cos \theta \, & , \; \;  y'(s) = \sin \theta \\
YI \theta'(s) = M \, & , \; \; M'(s) = N_x \sin \theta - N_y \cos \theta \\
N_x'(s) = - p^\mathrm{ext}_x \, & , \; \;  N_y'(s) = -p^\mathrm{ext}_y
\end{align}
\end{subequations}
where $s$ is the arc-length of the beam, $M$ the internal bending moment, $N=(N_x,N_y)$ the internal force, $(x,y)$ the current position of the central line, and $\theta$ the angle between the tangent of the centre line and the horizontal axis. The bending moment is linearly related to the curvature $\theta'(s)$ through the bending rigidity $YI$, where $Y$ is Young's modulus and $I$ is the second moment of area of the beam cross section. In the case of a rectangular cross section of thickness $h$ and width $w$, $I=h^3 w /12$ when bending occurs in the plane of the thickness $h$. The beam is clamped at $s=0$ and a vertical force $(0,-F)$ is applied at the $s=L$ extremity. We also consider the self-weight of the beam $p^\mathrm{ext}=(0,-\tau g )$, where $\tau$ is the mass per unit arc-length of the beam. The left boundary conditions are $x(0)=0$, $y(0)=0$, $\theta(0)=0$, and the right boundary conditions are $N_x(L)=0$, $N_y(L)=-F$, and $M(L)=0$.

The equilibrium equations can be recovered by considering the energy:
\begin{align}
E &=
\frac{1}{2} YI \int_{0}^{L} \left[ \theta' \right]^2 \d s 
+ \tau g \int_{0}^{L} y \, \d s 
+ F y(L)
\label{equa:energy-PE}
\end{align}
and the Lagrangian:
\begin{align}
{\cal L}(x,y,\theta) =
E  + \int_{0}^{L} \nu_i(s) \,  \left[x'-\cos \theta \right] \d s
 + \int_{0}^{L} \lambda_i(s) \,  \left[y'-\sin \theta \right] \d s
\label{equa:lagrangian-PE}
\end{align}
subjected to the left boundary conditions. The conditions for the vanishing of the first variation of the Lagrangian will yield the equilibrium equations
(\ref{sys:planar-elastica}) together with the right boundary conditions \citep{Audoly-Pomeau:Elasticity-and-Geometry:-From:2010}.

%\clearpage

\bibliographystyle{procRA} 
\bibliography{menisque}

\end{document}